# Proper Weyl Collineations in Kantowski-Sachs and Bianchi Type III Space-Times

Ghulam Shabbir

Faculty of Engineering Sciences

GIK Institute of Engineering Sciences and Technology

Topi Swabi, NWFP, Pakistan

Email: shabbir@giki.edu.pk

## Abstract

A study of proper Weyl collineations in Kantowski-Sachs and Bianchi type III space-times is given by using the rank of the $6 \times 6$ Weyl matrix and direct integration techniques. Studying proper Weyl collineations in each of the above space-times, it is shown that there exists no such possibility when the above space-times admit proper Weyl collineations.

## 1 Introduction

The aim of this paper is to study proper Weyl collineations (WCS) in Kantowski-Sachs and Bianchi type-III space-times by using the rank of the $6 \times 6$ Weyl metric and direct integration techniques. Through out $M$ denotes a (4-dimensional Connected, Hausdorff) smooth space-time manifold with Lorentz metric g of signature (-,+,+,+). The usual covariant, partial and Lie derivatives are denoted by a semicolon, a comma and the symbol $L$, respectively. The curvature tensor associated with $g_{ab}$, through the Levi-Civita connection, is denoted in component form where $R_{abcd}$, the Ricci tensor components are $R_{ab} = R^c{}_{acb}$, the Weyl tensor components are $C^a{}_{bcd}$, and the Ricci scalar is $R = g^{ab} R_{ab}$. Round and square brackets denote the usual symmetrization and skew-symmetrization.

Let $X$ be a smooth vector field on $M$ then in any coordinate system on $M$, one may decompose $X$ in the form

$$X_{a;b} = \frac{1}{2} h_{ab} + F_{ab} \tag{1}$$

where $h_{ab} = L_X g_{ab}$ and $F_{ab}(=-F_{ba})$ are symmetric and skew symmetric tensor on $M$, respectively. If $h_{ab} = f g_{ab}$ and $f(f : M \to R)$ is a real valued function on $M$ then $X$ is called a conformal vector field where $F_{ab}$ is called the conformal bivector. The vector field $X$ is called a proper conformal vector field if $f$ is not constant on $M$. For a conformal bivector $F_{ab}$ one can show that [1]

$$F_{ab;c} = R_{abcd} X^d - 2 f_{;[a} g_{b]c} \tag{2}$$

and

$$f_{a;b} = -\frac{1}{2} L_{ab;c} X^c - f L_{ab} + R_{c(a} F_{b)}{}^c \tag{3}$$

where $L_{ab} = R_{ab} - (1/6) R g_{ab}$. If $X$ is a conformal vector field on $M$ then by using (3) one can show that

$$L_X R_{ab} = -2 f_{a;b} - (f^c{}_{;c}) g_{ab}.$$

Further, the conformal vector field $X$ also satisfies [3]

$$L_X C^a{}_{bcd} = 0 \tag{4}$$

which can be written equivalently as

$$C^a{}_{bcd;f} X^f + C^a{}_{bcf} X^f{}_{;d} + C^a{}_{bfd} X^f{}_{;c} + C^a{}_{fcd} X^f{}_{;b} - C^f{}_{bcd} X^a{}_{;f} = 0.$$

The vector field $X$ satisfying the above equation is called a Weyl collineation (WC). The vector field X is called a proper WC if it is not conformal [2]. The vector field $X$ is called a homothetic vector field if $f$ is constant and a proper homothetic vector field if $f = \text{constan} t \neq 0$. If $f = 0$ on $M$ then vector field X is called a Killing vector field.

## 2 Main Results

It has been shown [2,4,5] that much information on the solutions of (4) can be obtained without integrating it directly. To see this let $p \in M$ and consider the following algebraic classification of the Weyl tensor as a linear map $\beta$ from the vector space of bivectors to itself; $\beta : F_{ab} \to F_{cd} C^{cd}{}_{ab}$, for any bivector $F_{ab}$ at $p$. The range of the Weyl tensor at $p$ is then the range of $\beta$ at $p$ and its dimension is the Weyl rank at $p$. It follows from [4] that the rank of the $6\times 6$ Weyl matrix is always even i.e. 6, 4, 2 or 0. If the rank of the $6\times 6$ Weyl matrix is 6 or 4 then every Weyl symmetry is a conformal symmetry [4,5]. For finding proper WCS, we restrict attention to those cases of rank 2 or less.

## 2.1 Proper WCS in Bianchi type III space-times

Consider a Bianchi type III space-time in the usual coordinate system $(t, r, \theta, \phi)$ (labeled by $(x^0, x^1, x^2, x^3)$, respectively) with line element [6,8]

$$ds^2 = -dt^2 + A(t)dx^2 + B(t)(d\theta^2 + \sinh^2\theta \, d\phi^2), \qquad (5)$$

where $A(t)$ and $B(t)$ are no where zero function of $t$. The non-zero independent components of Weyl tensor are

$$C_{0101} = \frac{1}{12AB^2}K(t) \equiv F1, \qquad C_{0202} = -\frac{1}{24A^2B}K(t) \equiv F2,$$

$$C_{0303} = \sinh^2\theta \; F2 \equiv F3, \qquad C_{1212} = \frac{1}{24AB}K(t) \equiv F4, \qquad (6)$$

$$C_{1313} = \sinh^2\theta \; F4 \equiv F5, \qquad C_{2323} = -\frac{\sinh^2\theta}{12A^2}K(t) \equiv F6,$$

where $K(t) = (B^2(-2\ddot{A}A + \dot{A}^2) + AB\dot{A}\dot{B} + 2A^2(\ddot{B}B - \dot{B}^2) + 4A^2B)$ and dot denotes the derivative with respect to $t$. The Weyl tensor of $M$ can be described by components $C_{abcd}$ written in a well known way [7]

$$C_{abcd} = diag(F1, F2, F3, F4, F5, F6).$$

We restrict attention to those cases of rank 2 or less, since by theorem [4] no proper WCS can exist when the rank of the $6\times 6$ Weyl matrix is $>2$. For the rank less or equal to two one may set four components of Weyl tensor in (6) to be zero. One gets $A$ and $B$ to be zero which gives contradiction to our assumption that $A$ and $B$ are no where zero functions on $M$ this implies that there exists no such possibility when the rank of the $6\times 6$ Weyl matrix is less or equal to zero. Hence no proper Weyl collineations exist in the above space-time (5).

## 2.2 Proper WCS in Kantowski-Sachs space-times

Consider a Kantowski-Sachs space-time in the usual coordinate system $(t,r,\theta,\phi)$ (labeled by $(x^0, x^1, x^2, x^3)$, respectively) with line element [6,8]

$$ds^2 = -dt^2 + A(t)dx^2 + B(t)(d\theta^2 + \sin^2\theta\, d\phi^2), \qquad (7)$$

where $A(t)$ and $B(t)$ are no where zero function of $t$. The non-zero independent components of Weyl tensor are

$$C_{0101} = \frac{1}{12AB^2} K(t) \equiv E1, \qquad C_{0202} = -\frac{1}{24A^2 B} K(t) \equiv E2,$$

$$C_{0303} = \sin^2\theta\, E2 \equiv E3, \qquad C_{1212} = \frac{1}{24AB} K(t) \equiv E4, \qquad (8)$$

$$C_{1313} = \sin^2\theta\, E4 \equiv E5, \qquad C_{2323} = -\frac{\sin^2\theta}{12A^2} K(t) \equiv E6,$$

where $K(t) = (B^2(-2\ddot{A}A + \dot{A}^2) + AB\dot{A}\dot{B} + 2A^2(\ddot{B}B - \dot{B}^2) + 4A^2 B)$ and dot denotes the derivative with respect to $t$. The Weyl tensor of $M$ can be described by components $C_{abcd}$ written in a well known way [7]

$$C_{abcd} = diag(E1, E2, E3, E4, E5, E6).$$

Here we will restrict our attention to those cases when the rank of the $6\times 6$ Weyl matrix is 2 or less, since by theorem [4] no proper WCS can exist when the rank of the $6\times 6$ Weyl matrix is $> 2$. For the rank less or equal to two one may set four components of Weyl tensor in (8) to be zero. One gets $A$ and $B$ to be zero which gives contradiction to our assumption that $A$ and $B$ are no where zero functions on $M$ this implies that there exists no possibility when the rank of the $6\times 6$ Weyl matrix is less or equal to zero. Hence again no proper Weyl collineations exist in the above space-time (7).

## Summary

In this paper a study of proper Weyl collineations in Kantowski-Sachs and Bianchi type III space-times is given by using the rank of the $6\times 6$ Weyl matrix and direct integration techniques and the theorem given in [4]. Studying proper Weyl collineations in the Kantowski-Sachs and Bianchi type III space-times, it is shown that the above space-times do not admit proper Weyl collineations.

# References

[1]     G. S. Hall, Gen. Rel. Grav., **22** (1990) 203.

[2]     G. Shabbir, Ph. D. Thesis University of Aberdeen, 2001.

[3]     G. S. Hall, "Proceedings of The Hungarian Relativity Workshops", Tihany, Hungary, 1989.

[4]     G. S. Hall, Gravitation and Cosmology, **2** (1996) 270.

[5]     G. S. Hall, Curvature and Physics, Kazan, 1998.

[6]     H. Baofa, Int. J. Theor. Phys., **30** (1991) 1121.

[7]     G. Shabbir, Class. Quantum Grav., **21** (2004) 339.

[8]     J. D. Lorenz, Phys. A, **15** (1982) 2809.